# Lattice effect on electric and magnetic resonance overlap in periodic array

Viktoriia E. Babicheva* and Jerome V. Moloney

College of Optical Sciences, University of Arizona,
1630 E. University Blvd., P.O. Box 210094, Tucson, AZ 85721

vbab.dtu@gmail.com

**Abstract.** Designing the shape of silicon nanoparticles has been shown to be an effective approach to increasing overlap between electric and magnetic dipole resonances thereby achieving directional scattering and decrease of reflection. Variations of disk diameter and/or height affect resonances differently and can thus result in resonance overlap. In most of the studies, the disks are arranged in a periodic array where the periodicity is varied together with disk diameter, but the role of lattice effect is neglected. Here we theoretically study a periodic array of disks and show that the contribution of the lattice effect in shifting resonance positions is comparable to the effect of the diameter change. We demonstrate that the lattice effect is important even when the wavelength of diffraction remains on the blue side from electric and magnetic dipole resonances and there are no additional lattice resonances are excited. Period and disk dimensions are chosen so that the resonances overlap in the proximity of the telecommunication wavelength which is of great practical interest.

Subwavelength nanostructures have been actively studied with the aim of being applied in optical and photonic devices, for sensing [1], photovoltaics [2,3], optical communications [4], detectors [5], functional elements, including metasurfaces [6], directional scattering and enhanced light-matter interaction [7], high-resolution near-field microscopy [8,9], and others. Various nanoparticle shapes [10-12] and materials [13-15] have been proposed to serve as building blocks to enhance resonances and increase scattering and/or absorption. Recently, high-index dielectrics, such as silicon, germanium, and III-V compounds, have been shown to be a promising platform for realizing subwavelength nanoantennas. The operation bandwidth of such nanostructures is mainly defined by the material optical properties and particle dimensions. Larger or smaller particles can shift resonances to larger or smaller wavelength, correspondently. Thus, with the silicon particles, the effect can be achieved in mid-infrared or visible spectral range. However, material losses may hinder nanostructure performance, and in general, it is desirable to operate at the energies below semiconductor bandgap (e.g. 1.1 eV = 1.13 μm for silicon).

One of the advantages of high-index antennas over conventional plasmonic particles is the possibility to excite both electric and magnetic resonances in particles of simple shape, like spheres or disks [16-18]. The strong electric and magnetic resonance excitations and their interference in high-index nanostructures open a unique opportunity for designing optical nanoantennas and metasurfaces with a considerable enhancement of the nonlinear response [19]. While silicon possesses an almost negligible second-order nonlinearity of bulk material, other semiconductors such as aluminum gallium arsenides can bring a significant enhancement. It drives a rapid progress in developing nonlinear optical nanostructures beyond the diffraction limit and useful applications in nonlinear photonic metadevices operating at the nanoscale [20,21]. A dynamic control of optical response of semiconductor nanoparticle arrays can be achieved either by combining them with elements of other materials with the strong dynamic response, such as vanadium dioxide [22] or ferroelectrics [23], or by inducing a switch in the semiconductor itself. For example, semiconductor nanoparticles, including silicon and III-V compounds, designed as nonlinear resonators can be utilized for ultrafast switching [24]. The transient modulation of the material permittivity by variations of electron-hole plasma density induced via its photoexcitation by femtoseconds laser has been shown significantly alters scattering behavior of silicon nanoparticles [25].

The interplay of resonances, for instance, their overlap, offers the potential to tune scattering and engineer more efficient subwavelength antennas. The Kerker effect is defined as near-zero backward scattering when electric (ED) and magnetic dipole (MD) moments of the particle are equal in magnitude and in phase [26-28]. In a similar way, one can achieve a strong backward reflection in the structure where ED and MD moments are out of phase. One can utilize such nanoparticle array to compensate reflection from the high-index substrate, and in this case, the array serves as an antireflective coating [2,29].

One of the practical interests is to achieve the Kerker effect when multipole moments are in resonant states, and for instance, ED and MD resonances overlap (EDR and MDR, respectively). For silicon spherical particles this is not possible, but adjustment of diameter and/or height of disk particles or designing particle shape can bring resonances into overlap [11,30]. The effect of disk height with fixed diameter has been studied for a single silicon particle in detail proving the EDR and MDR shift differently upon height changes [30]. The seminal work [6] demonstrates the EDR and MDR overlap in the array of disks, and this overlap is attributed to the changes in the disk diameter. However, together with the disk dimensions, the distance between disks in the array is also varied, and the effect of these changes has remained overlooked in [6]. As we show in the present work, nanoparticle resonances are strongly affected by their arrangement even in case of subdiffraction distances, and this effect can play a crucial role in defining the resonance positions and their overlap.

Arranging particles in an array adds new features to the spectrum of optical response as the contribution of all particles in the lattice affects the resonance position. Rectangular lattices, where the particles are periodic and equally spaced in mutually perpendicular directions (e.g. periods $D_x$ and $D_y$ in x- and y-directions, Fig. 1a), have been studied in detail. The best known and studied effect is lattice resonance [1,3,13,31-41], which is an additional narrow resonance in the proximity to the wavelength of diffraction, the so-called Rayleigh anomaly. Usually, the most pronounced lattice resonance is excited when the Rayleigh anomaly is on the red side from the single-particle resonance. It has been shown that the lattice resonance can be an excitation of the ED or MD [13] as well as electric quadrupole [35,38], and by analogy, one can expect that higher multipoles can also contribute to the lattice resonance [42].

In the present work, we consider only dipole moments of the disk particle as higher multipole contributions in the scattering properties of the silicon disk with such dimensions are negligible in comparison to dipole contributions (see calculations of the different multipole moments for the similar size disk in [43]). For a dipole array, the lattice effect is defined by the periodicity in the direction of dipole emission. It has been shown previously that when an electric field of the incident plane wave is polarized along the x-axis, the array period in the $D_x$ direction affects the MD moment and is responsible for the excitation of an additional MD lattice resonance [13]. Correspondingly, in the same polarization, tuning the period $D_y$ affects the ED moment and is responsible for an ED lattice resonance (Fig. 1a). Thus ED and MD lattice resonances can be independently controlled by varying the mutually-perpendicular array periods, and the lattice resonances can be brought into overlap without imposing limitations on particle shape [37,44].

As the lattice resonance is typically narrow and appears as an additional resonance in the proximity to the Rayleigh anomaly wavelength, the lattice resonance is usually easy to recognize. However, much less attention has been paid to another signature of the lattice effect, such as a change of resonance profile and spectral position when the Rayleigh anomaly is on the blue side from the single-particle resonance and no additional lattice resonance is excited. Furthermore, there is a misconception that lattice effect can be neglected once the single-particle resonance is on the red side from Rayleigh anomaly [6,43,45]. This assumption is not valid, and here we show that taking into account array periodicity is important even if particle resonances are spectrally separated from Rayleigh anomaly and the latter one is on a blue side from the resonance.

In the present work, we report on the theoretical observation of the lattice effect in a disk array, and we directly compare the lattice effect on the EDR and MDR shifts to the adjustment of disk diameter. We demonstrate that the array period and disk diameter have a comparable contribution to the resonance positions. We choose array parameters the same as in the experimental work [6], and EDR and MDR overlap in the proximity of the telecommunication wavelength. We show that the EDR and MDR overlap should be attributed to not only change of particle shape but also to the array periodicity.

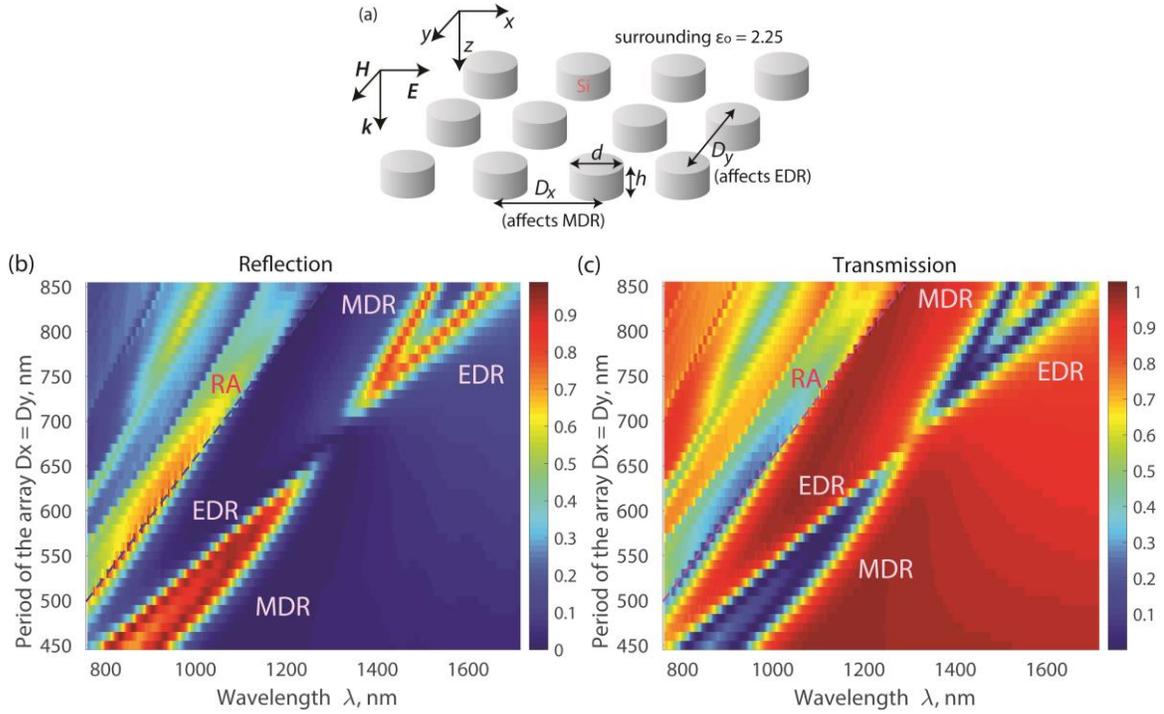

Fig. 1. (a) Schematics of the structure under consideration: periodic array of disks and normal incidence of the light with the E field along the x-axis. Surrounding medium has $\varepsilon_o$ = 2.25. For this polarization, the array period in the x-direction $D_x$ affects excitation of a magnetic dipole resonance (MDR), and the period in y-direction $D_y$ affects the electric dipole resonance (EDR). (b) Reflection and (c) transmission through the disk array under variation of the array periods and disk diameters with $D_x = D_y = d + 200$ nm. The dashed blue and red lines mark the Rayleigh anomaly wavelength $\lambda_{RA} = D_x = D_y$.

We study a periodic array of disks with permittivity $\varepsilon$ = 12.25 + 0.05i (silicon), diameter d, fixed height h = 220 nm, arranged in the array with periods $D_x$ and $D_y$ and surrounded by a medium with fixed permittivity $\varepsilon_o$ = 2.25 (Fig. 1a). Disk shape has a particular advantage from the practical point of view: in contrast to elliptical particles, it can be straightforwardly fabricated with lithographical methods and in contrast to cuboids, disks do not have sharp edges in an in-plane direction that require additional precision in fabrication. The disk array can be fabricated by structuring silicon layer on a fused silica substrate via an electron-beam lithography-based process (for the similar arrays, the details of structure fabrication and characterization can be found in [6,43,45]). We perform numerical simulations with the finite-difference time-domain (FDTD) method with an in-house package. The unit cell contains one disk, periodic boundary conditions in the x- and y-directions, perfectly matched layers in the z-direction, and light is incident normal to the array with the E field along the x-axis.

To start with, we perform a study similar to the one in [6]: we vary simultaneously d, $D_x$, and $D_y$ satisfying the condition $D_x = D_y = d + 200$ nm and calculate reflection from the array (Fig. 1b). In agreement with Ref. [6] results for the disk array, we see that under the condition d = 500 nm, EDR and MDR spectrally overlap and reflection is suppressed to a near-zero value. Transmission calculations complement the reflection map and confirm resonance overlap and increased transmission for d = 500 nm (Fig. 1c).

One can compare particle resonances in the array to the single particle resonances. In our case, resonances of the single silicon disk (without the array) with the same height but varying diameter are calculated in [6] and indicate that EDR and MDR overlap occurs for d = 290 nm at the wavelength about 1050 nm. The calculations also show that for the single disk with d = 500 nm, MDR and EDR positions are $\lambda_{MDR} \approx 1350$ nm and $\lambda_{EDR} \approx 1600$ nm, correspondingly. Obviously, the optimum disk diameter and wavelength of the resonances calculated for single disk are very different from the parameters calculated for the disk array, in both Ref. [6] and the present work (Fig. 1b,c). This discrepancy is caused by the lattice effect of compactly arranged particles in the array, and the lattice contribution cannot be neglected in this case.

As the arrangement of particles in the array plays a crucial role in the position of particle resonances, let us consider what structural parameters affect the disk resonances. Resonance overlap and reflection suppression in Fig. 1 is a typical manifestation of Kerker effect, however, the simultaneous changes of d, $D_x$, and $D_y$ cause an ambiguity in identifying the main cause of EDR and MDR spectral shifts. The work [6] ascribes the effect of resonance overlap solely to the changes in disk diameter. The change of EDR and MDR positions are described as if they were excited in a single particle. However, along with the diameter change, the array periods change as well, and below we show that this period variation strongly affects the resonance position.

Let us analyze what effect causes changes in the array period rather than the effect of disk diameter. Figure 2a shows calculations for $D_y$ = 700 nm, d = 500 nm, and different $D_x$. One can see that position of EDR is fixed to $\lambda \approx$ 1350 nm and MDR changes position from $\lambda_{550} \approx 1180$ nm to $\lambda_{850} \approx 1450$ nm remaining parallel to the Rayleigh anomaly. Similar to Fig. 1b, the resonance overlap causes a reflection suppression which means the conditions for the Kerker effect are satisfied. As the resonance position is defined by the lattice period, either EDR or MDR can be excited at a larger wavelength (either $\lambda_{EDR} > \lambda_{MDR}$ or $\lambda_{MDR} > \lambda_{EDR}$) depending on the particular period values, so the order of resonances in Fig. 2a is different from the one in Fig. 1b. Both EDR and MDR maintain a spectral separation $\Delta\lambda = \lambda_{EDR(MDR)} - \lambda_{RA-EDR(MDR)} \approx 200$ nm from the position of Rayleigh anomaly that they correspond to: $\lambda_{RA-EDR} = D_y = 700$ nm and $\lambda_{RA-MDR} = D_x$. In Fig. 2a, we see that at the large $D_x$, the MDR is closer to Rayleigh anomaly than for the small $D_x$, and it is a general property of the lattice effect in the nanoparticle array: the sparser the array the closer the resonance to Rayleigh anomaly wavelength. Similar to the effect of $D_x$ changes and MDR shift, the orthogonal period $D_y$ variation causes an EDR shift (Fig. 2b). In both panels of Fig. 2, the disk diameter is fixed to d = 500 nm, and the calculation results indicate that resonance position is strongly affected by the changes of array period even for the fixed disk size. The results in Fig. 1 and work [6] cannot be ascribed solely to the effect of diameter change and needs to account for the array periodicity.

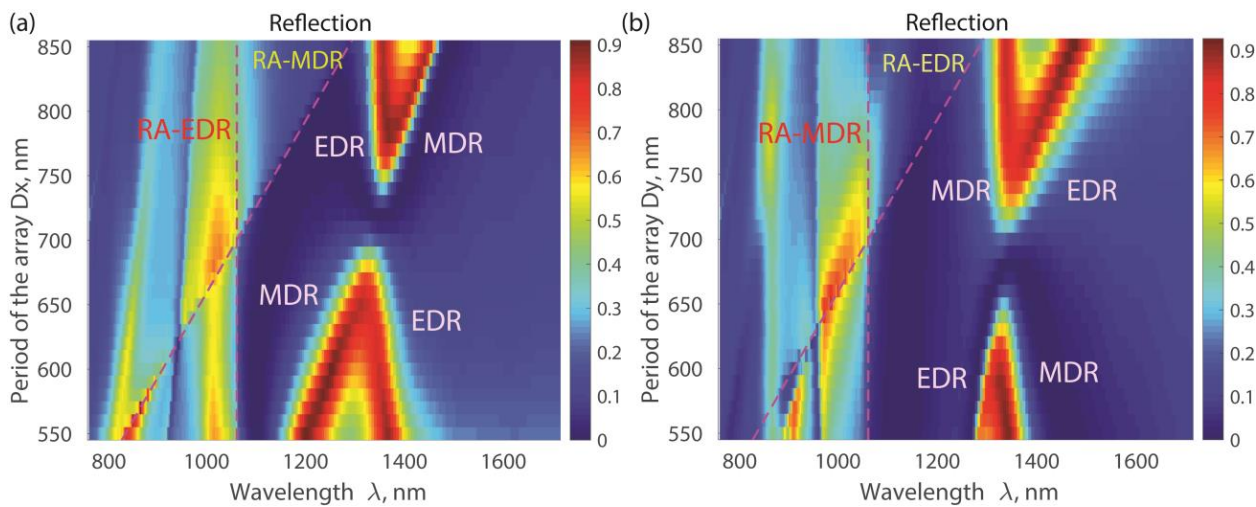

Fig. 2. Reflection of disk array under independent variations of (a) $D_x$ and (b) $D_y$. The second period is fixed at 700 nm, and the diameter d = 500 nm is fixed and the same on both panels. In (a), $D_x$ changes cause MDR shift, but $D_y$ and EDR position remain unchanged. In (b), $D_y$ changes cause EDR shift, but $D_x$ and MDR position remain unchanged. The dashed magenta lines mark the Rayleigh anomaly wavelength: $\lambda_{RA-EDR} = D_y = 700$ nm and $\lambda_{RA-MDR}$

= $D_x$ in (a), and $\lambda_{RA-MDR}$ = $D_x$ = 700 nm and $\lambda_{RA-EDR}$ = $D_y$ in (b). Both EDR and MDR remain on spectral distance about $\Delta\lambda \approx$ 200 - 350 nm from corresponding Rayleigh anomaly wavelength.

Finally, let us consider separate changes of lattice period and disk diameter in the structure that follow the initial case of geometry with $D_x$ = $D_y$ = d + 200 nm. To differentiate the effects of lattice and disk diameter changes and compare contributions from each of them on the resonance position, we separately change $D_x$, $D_y$, and d and analyze changes in reflection profiles (Fig. 3). We choose two sets $D_x$ = 450 nm and $D_x$ = 500 nm, and both sets satisfy the initial condition $D_x$ = $D_y$ = d + 200 nm. For the intermediate step $D_x$ = 500 nm, $D_y$ = 450 nm, and d = 250 nm, we see that the MDR shift due to $D_x$ change from 450 nm to 500 nm (black to cyan line) is comparable to the effect of d change from 250 nm to 300 nm (cyan to blue line, Fig. 3a). Similarly, for the intermediate step $D_x$ = 450 nm, $D_y$ = 500 nm, and d = 250 nm, the calculations show that the EDR shift due to $D_y$ change from 450 nm to 500 nm (black to red line) is comparable to the effect of the d change from 250 nm to 300 nm (red to blue line, Fig. 3b). Figure 3 shows calculations only for two sets of parameters, but the conclusion is valid for larger periods and disk diameters, and the lattice effect should be taken into account in such particle arrays.

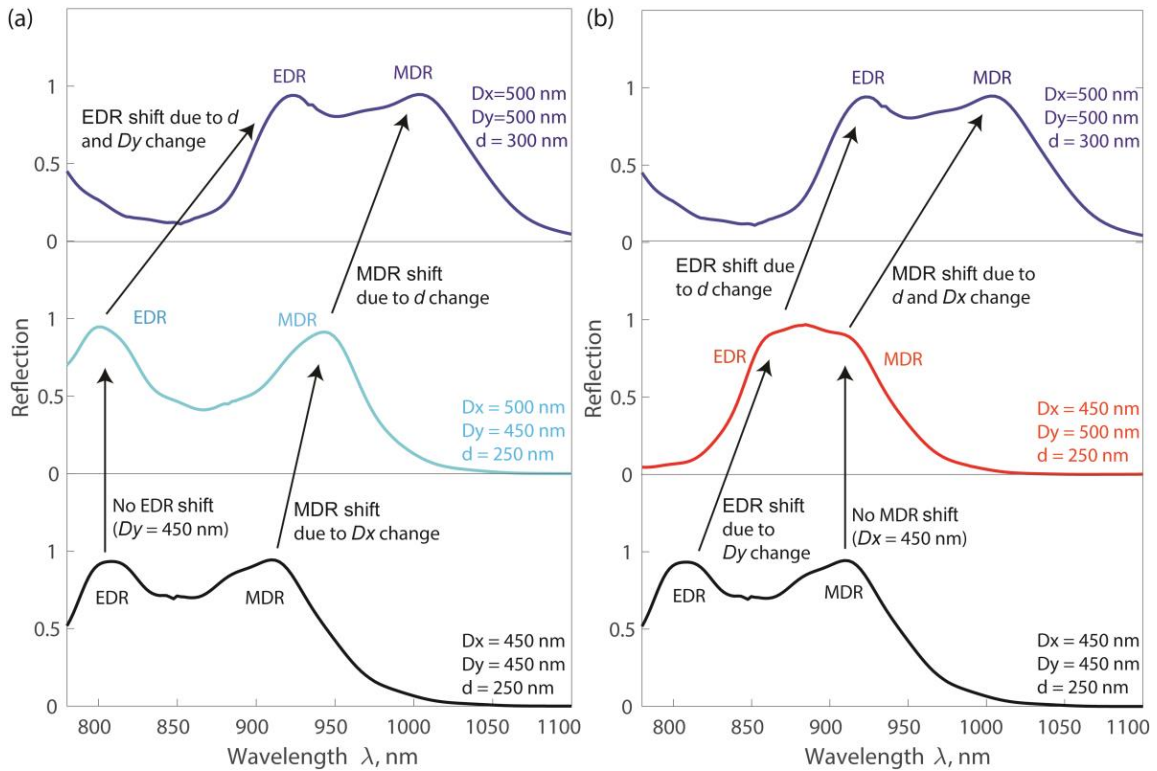

Fig. 3. Contrast of lattice and disk diameter variation influence: consequent changes of $D_x$, $D_y$, and d between the cases $D_x$ = $D_y$ = d + 200 nm = 450 nm and $D_x$ = $D_y$ = d + 200 nm = 500 nm. (a) The intermediate step is $D_x$ = 500 nm, $D_y$ = 450 nm, and d = 250 nm, and the MDR shift caused by $D_x$ change (black to cyan line) is comparable to the effect of d change (cyan to blue line). (b) The intermediate step is $D_x$ = 450 nm, $D_y$ = 500 nm, and d = 250 nm, and the EDR shift caused by $D_y$ change (black to red line) is comparable to the effect of d change (red to blue line).

To sum up, we have considered a silicon particle array which is optimized to increase transmission and resonantly scatter light in the forward direction in the telecommunications wavelength range. In particular, the reflection from the array is nearly zero and transmission is almost 100% at the wavelength 1300 nm, which is achieved for the array with silicon disk particles of 470 nm in diameter and 220 nm in height arranged into the square array with the period of 670 nm. We have studied a periodic array of disks with various dimensions and analyzed the effect of lattice period and disk diameter on the EDR and MDR position and their overlap. We have shown that resonance position can be changed by the period, and EDR and MDR can be brought into overlap solely by the change of array period having disk diameter fixed. We emphasize that the Rayleigh anomaly remains on the blue side from the resonance and no additional resonances (so-called lattices resonances) are excited in the structure. The lattice effect plays a crucial role in this structure, and EDR and MDR closely follow the corresponding Rayleigh anomaly and mainly defined by its position even though the resonances are about $\Delta\lambda \approx 200 - 350$ nm away from Rayleigh anomaly. We have also shown that the lattice effect is comparable to the effect of disk size change, and thus cannot be neglected in the closely spaced particle arrays despite the resonances excited at the wavelength larger than Rayleigh anomaly.

**Acknowledgment**

The authors thank John Nehls and Colm Dineen for the help with cluster machine. This material is based upon work supported by the Air Force Office of Scientific Research under Grant No. FA9550-16-1-0088.